\documentclass[preprint]{aastex}
\let\captionbox\relax
\usepackage{natbib}
\usepackage{morefloats}
\usepackage{amsmath}
\usepackage{graphicx}
\usepackage{verbatim}
\usepackage{color}
\usepackage{ulem}
\usepackage{appendix}
\usepackage{hyperref}
\usepackage{caption}
\usepackage{subcaption}
\usepackage{calc}
\begin{document}
\newsavebox\CBox
\newcommand{\fig}[1]{Figure \ref{#1}}
\newcommand{\eq}[1]{equation (\ref{#1})}
\newcommand{\eqs}[2]{equations (\ref{#1}) - (\ref{#2})}
\newcommand{\vc}[1]{ {\bf{#1}} }
\newcommand{\grad}[1]{ \vc{\nabla } #1 }
\newcommand{\dv}[1]{ \vc{\nabla } \cdot #1 }
\newcommand{\dt}[1]{ \frac{\partial}{\partial t} #1 }
\newcommand{\dx}[2]{ \frac{\partial}{\partial #1} #2 }
\newcommand{\rev}[1]{#1}
\newcommand{\rthree}[1]{#1}

\title{A Very Fast And Angular Momentum Conserving Tree Code}
\author{Dominic C. Marcello}	
\email{dmarce504@gmail.com}

\affil{Department of Physics \& Astronomy}
\affil{Louisiana State University}
\affil{Baton Rouge, Louisiana}

\begin{abstract}{There are many methods used to compute the classical gravitational field in astrophysical simulation codes. With the exception of the typically impractical method of direct computation, 
none ensure conservation of angular momentum to machine precision. Under uniform time-stepping, the Cartesian fast multipole method of Dehnen (also known as the very fast tree code) conserves linear
 momentum to machine precision. We show it is possible to modify this method in a way that conserves both angular and linear momenta.}
\end{abstract}
\section{Introduction}

Angular momentum  plays an important role in a plethora of astrophysical phenomena. These phenomena include, but are not limited to, the formation of galactic discs, the accretion of matter in interacting 
binary star systems, the formation of proto-stars and proto-planetary discs, the dynamics of planetary orbits, and rapidly rotating neutron stars and black holes. In the limit that Newtonian gravity  provides
an accurate description of the gravitational field of any isolated astrophysical system, we expect the total angular momentum change due to the gravitational interaction to be zero. 

Classical self gravitating astrophysical simulation codes compute the gravitational field in a variety of ways. 
 Grid based hydrodynamics codes 
\citep[e.g.][]{FORTZLMRTT2000,
ABBDHJLNSZ2010,
STGTHS2008,
DL2012,
DMTF2006} 
 solve the discretized Poisson's equation for the gravitational potential using iterative techniques, Fourier transforms, or a combination of the two. N-body codes and smoothed particle hydrodynamics (SPH) codes 
\citep[e.g.][]{HK1989,
S2005, 
VKPB2009, 
LIG2010,
YB2012}
may use the tree code of \citet{BH1986}, particle-mesh methods \citep{B2002}, or fast multipole methods (FMM) \citep{GR1997, WS1995, D2000, DR2011}. Some of the gravity solvers in the aforementioned codes conserve linear momentum, 
but none ensure conservation of angular momentum.

\rev{Due to the symmetry of the equations for the multi-pole interactions, the method of \citet{D2000} (hereafter ``D2000'') naturally conserves linear momentum between any pair of particles. We have developed
a modification to D2000 that preserves this property, while simultaneously conserving  angular momentum between any pair of interacting {\it multi-poles}. This kind of conservation is not of the same
quality as the conservation of linear momentum. It introduces artificial torques, however, the added torques are within the error bound of the original scheme.  }
\rev{In \S{\ref{Method}} we describe our modification to D2000.} In \S{\ref{Numerical}} we provide a numerical test of the method. \rev{In \S {\ref{discuss}} we make the case for using this technique to
model double white dwarfs (DWDs), as well as discuss some of the method's shortcomings.}
In \S A we provide a more general derivation of the method that applies to higher orders.

\section{Method}
\label{Method}

The algorithm presented by D2000 decomposes the set of particles into an oct-tree structure, with each cell in the oct-tree containing a predetermined maximum number of particles, $n_\mathrm{crit}$. The
code presented in this paper was run with \rev{$n_\mathrm{crit} = 25$}. Two cells, cell ``A" and cell ``B", are considered
``well separated" if they satisfy the ``opening criterion",
\begin{equation}
|\vc{Z_A} - \vc{Z_B}| \geq  \frac{1}{\theta} \left(R_{A,\mathrm{max}} + R_{B,\mathrm{max}}\right) ,
\end{equation}
where  $\vc{Z_A}$ and $\vc{Z_B}$ are the respective centers of mass of cells A and B, $R_{A,\mathrm{max}}$ and  $R_{B,\mathrm{max}}$ are the maximum distances from a particle within the cells to the centers of mass of their respective
cells, and $\theta$ is an adjustable parameter called the ``opening angle", where $0 < \theta \leq 1$. Forces within a cell are computed using multipole interactions and Taylor expansions for all cells
that are well separated from it, while the force contributions from any remaining nearby particles are computed directly. 
(Note that \citet{D2014} has recently developed a more complex selection criteria that uses an error estimate to select interaction pairs in a manner that maximizes execution speed for a given error. The development we present here is also applicable to that method.) 
Here we will present only what is necessary to describe the modifications we have made. 
Refer to D2000 for the full description of the original method.

Let $\vc{R} = \vc{Z_B} - \vc{Z_A}$. Within cell A there are $N_A$ particles located 
at positions $\vc{X_1}, \vc{X_2}, ..., \vc{X_{N_A}}$ and with masses  $\mu_{A 1}, \mu_{A 2}, ..., \mu_{A N_A}$. Similarly,  within cell B there are $N_B$ bodies located 
at positions $\vc{Y_1}, \vc{Y_2}, ..., \vc{Y_{N_B}}$ and with masses  $\mu_{B 1}, \mu_{B 2}, ..., \mu_{B N_B}$. Let $\vc{x_n} = \vc{X_n} - \vc{Z_A}$ and $\vc{y_n} = \vc{Y_n} - \vc{Z_B}$. The 
monopole\rev{, dipole, and quadrupole} moments are
\begin{equation}
\label{eq1}
M := \sum_{\vc{X_n}  \in \mathrm{cell}} \mu_n,
\end{equation}
\begin{equation}
M_{i} :=\sum_{\vc{X_n} \in \mathrm{cell}}  \mu_{n} x_{n,i},
\end{equation}
\rev{and}
\begin{equation}
\label{quad}
M_{i j} :=\sum_{\vc{X_n} \in \mathrm{cell}}  \mu_{n} x_{n,i} x_{n,j}.
\end{equation}
Gradients of the Green's function for the gravitational potential, $\vc{\nabla^n} g\left(R\right) = -\vc{\nabla^n} \frac{1}{R}$, are 
\begin{equation}
D := -\frac{1}{R},
\end{equation}
\begin{equation}
D_i := \frac{R_i}{R^3},
\end{equation}
\begin{equation}
D_{i j} := -\frac{3 R_i  R_j - \delta_{i j} R^2}{R^5},
\end{equation}
and
\begin{equation}
D_{i j k} := \frac{15 R_i R_j R_k - 3 \left(\delta_{i j} R_k + \delta_{j k} R_i + \delta_{k i} R_j\right) R^2}{R^7}.
\end{equation}
The approximated potential generated by the particles in cell B at a position $\vc{x} := x_i \vc{\hat{e}_i}$ in cell A is then
\begin{multline}
\label{eq2}
\Phi_{B\rightarrow A}\left(\vc{X}\right)  \approx \left( M_B D - M_{B i} D_i + \tfrac{1}{2} M_{B i j} D_{i j} \right) +  \\
                      x_i \left( M_B D_i - M_{B j} D_{i j} + \tfrac{1}{2} M_{B j k} D_{i j k}\right) +
     \tfrac{1}{2} x_i x_j \left(  M_B D_{i j} - M_{B k} D_{i j k} \right) +
   \tfrac{1}{6} x_i x_j x_k M_B D_{i j k}.
\end{multline} 
The Cartesian FMM described by \eqs{eq1}{eq2} is the same as in D2000, except that \rev{: (1)} we have opted to express the quadrupole moments in the extensive form \rev{, and (2) we have added terms that involve the dipole moment. These terms drop out in the case that cell coordinate centers coincide with cell centers of mass. }
As in D2000, we have dropped the octupole moment from \eq{eq2}. For a given interaction, this term is constant in space and hence does not contribute to the force calculation.

Using \eq{eq2}, the gravitational acceleration caused by the particles in cell B at a point $\vc{X}$ within cell A can be expressed as
\begin{multline}
\label{eq3}
\vc{g}_{B\rightarrow A}\left(\vc{X}\right) = -\left[\left( M_B D_i - M_{B j} D_{i j} + \tfrac{1}{2} M_{B j k} D_{i j k}\right) + \right. \\ 
\left. x_j \left(M_B D_{i j} -  M_{B k} D_{i j k} \right) +
\tfrac{1}{2} x_j x_k M_B D_{i j k} \right] \vc{\hat{e}_i}.
\end{multline}
Similarly,  the gravitational acceleration caused by the particles in cell A at a point $\vc{Y}$ within cell B can be expressed as
\begin{multline}
\label{eq4}
\vc{g}_{A\rightarrow B}\left(\vc{Y}\right)  = -\left[ \left(-M_A D_i - M_{A j} D_{i j} - \tfrac{1}{2} M_{A j k} D_{i j k}\right) + \right.\\ 
\left. y_j \left(M_A D_{i j} + M_{A k} D_{i j k} \right) -
 \tfrac{1}{2} y_j y_k M_A D_{i j k} \right] \vc{\hat{e}_i}.
\end{multline}
\rev{ Using equations (\ref{eq1}) - (\ref{quad}),  (\ref{eq3}), and (\ref{eq4}), we can express the force between two unit masses as 
\begin{equation}
\label{unit}
\vc{g}_{\mp \rightarrow \pm} = \mp \left(  D_i + \left( x_j - y_j\right) D_{i j} +  \left( x_j - y_j\right) \left( x_k - y_k\right) D_{i j k} \right) \vc{\hat{e}_i}.
\end{equation}
Although the computed force is an approximation of the force on an individual particle, \eq{unit} shows that the sum of forces between any two particles is exactly zero.  This implies the sum
of linear momentum changes due to gravitation over all the masses in pairs of interacting cells is zero, and therefore the change over the entire computational domain is zero.
}

The same result does not generally hold for the sum of the torques generated \rev{between pairs of cells}.
Referring to the more general derivation of the method in \S A, we write \eq{imbalance_eq} to expansion order $P=3$ and find the sum of all torques to be
\begin{equation}
\label{nettorque}
\vc{\tau_{A B}} = \tfrac{1}{2} \epsilon_{p i q} \left( M_{A p j k} M_B  - M_A M_{B p j k}\right)  D_{i j k} \vc{\hat{e}_q},
\end{equation}
where we define the octupole moments, 
\begin{equation}
M_{p j k} := \sum_{\vc{X_n} \in \mathrm{cell}}  \mu_{n} x_{n,p} x_{n,j} x_{n,k}.
\end{equation}
\rev{For many evolution methods employing the FMM, such as SPH or N-body, the net torque found in \eq{nettorque} (or the equivalent expression for a higher expansion order) is the sole source of angular momentum non-conservation. Eliminating
this efficiency would, therefore, guarantee angular momentum conservation to machine precision}.

We seek a correction to the Cartesian FMM of D2000 that (1) balances the net torque found in \eq{nettorque}, (2) produces an equal and opposite force on each cell, and (3) is within 
the error bounds of the computed force. 
\rev{One} possible solution satisfying these requirements uses the corrective force
\begin{equation}
\label{eq7}
\vc{F_c} = -\tfrac{1}{2} \left( M_{A j k l} M_B - M_A M_{B j k l}\right) D_{i j k l} \vc{\hat{e}_i},
\end{equation}
\rev{where $D_{i j k l}$ is the fourth derivative of the Green's function.}
Proof that $\vc{F_c}$ cancels the torque imbalance is found in \S A.
The correction for cell A is 
\begin{equation}
\vc{g}_{C, B\rightarrow A}\left(\vc{X}\right) = +\frac{1}{M_A} \vc{F_c}
\end{equation}
 and the correction for cell B is \begin{equation}
\vc{g}_{C, A\rightarrow B}\left(\vc{Y}\right) = -\frac{1}{M_B} \vc{F_c}.
\end{equation}
The corrective accelerations, $\vc{g}_{C, A\rightarrow B}$ and $\vc{g}_{C, B\rightarrow A}$, are added to $\vc{g}_{A\rightarrow B}$ and $\vc{g}_{B\rightarrow A}$, respectively, to obtain the total
acceleration. This correction produces an extra acceleration that is constant over each cell. The torque produced is equal in magnitude but opposite in direction to the
torque imbalance found in \eq{nettorque}.  The sum of the corrective forces on one cell is equal and opposite to that on the other cell, preserving the force balance of the original method. 
Because it uses a higher 
order Green's function derivative, the corrective force is within the error bounds of the original method.

Another possible solution is to replace $D_{i j k l}$ in \eq{eq7} with the non-symmetric tensor,
\begin{equation}
{D'}_{i j k l} := \\
+\frac{15 \left(\delta_{i j} R_k R_l + \delta_{i k} R_j R_l + \delta_{i l} R_j R_k \right)}{R^7}
\\
-\frac{3 \left( \delta_{ i j } \delta_{ k l } + \delta_{ i k } \delta_{ j l } + \delta_{ i l } \delta_{ j k } \right)}{R^5}, 
\end{equation}
yielding the alternative force correction, 
\begin{equation}
\label{eq8}
\vc{F'_c} = -\tfrac{1}{2} \left( M_{A j k l} M_B - M_A M_{B j k l}\right) {D'}_{i j k l} \vc{\hat{e}_i}.
\end{equation}
Note that ${D'}_{i j k l}$ is simply ${D}_{i j k l}$ with any terms that do not contribute to \eq{nettorque} removed. As shown below in  \S{\ref{Numerical}},
\eq{eq8} yields a faster implementation for a given opening angle while slightly increasing the solution error.

\rev{It is important to note that the quality of torque conservation in our modified FMM is not the same as the quality of force conservation. As shown by \eq{unit}, the force between any two {\it individual} particles 
sums to zero. An analogous relation does not hold for the torques. The torques satisfy the less strict requirement that the sum of torques between all the masses in two interacting cells is zero.
The correction also introduces unphysical torques between particles in the same cell, however, these corrections are within the error bounds of the original scheme.
}

\section{Numerical Test}
\label{Numerical}
To test our new method, we have implemented a minimalistic version of the method of D2000, with options to use the corrections described by equations (\ref{eq7}) or (\ref{eq8}). This code is written in C++ for serial execution 
\rev{on a single processing core}. 
 The version of the code used in this paper is available \rthree{ through the Zenodo repository, \\
 \rev{\url{https://doi.org/10.5281/zenodo.571523}.}} Note that the code is intended only to illustrate our method, and is not intended for production purposes.

For our test problem, we have chosen a binary star system for which the net torque imbalance can be relatively high.  As can be seen in \eq{nettorque}, the torque imbalance grows with the difference between octupole moments of the stellar components of a binary. Therefore, the larger and less centrally condensed one star is compared to its companion, the larger the net torque imbalance. \rthree{One such system is a high mass ratio double white dwarf (DWD) with the larger, less massive star filling its Roche lobe. A system like this, if  stable to mass transfer, is a potential progenitor of an AM Canum Venaticorum (AM CVn) type cataclysmic variable binary star \citep[][]{MNS2004,JBHGBA2016}. 
Our test problem is an approximation of such a system. The accretor has a mass of $1 M_\odot$ and the donor a mass of $0.2 M_\odot$, with the donor's volume equal to the volume of its Roche lobe. Each component is taken to be a spherical Lane-Emden polytrope. In realistic systems, the donor will be tidally distorted, however, the spherical approximation is sufficient to demonstrate the usefulness of our method. The donor has a polytropic index of $\frac{3}{2}$ and the accretor has a polytropic index of $3$, approximating the cold white dwarf equation of state in the low and high mass limits, respectively. The test problem consists of $10^6$ equal mass particles, chosen by sampling the density distribution computed from integrating the Lane-Emden equation for each component. }

Our \rev{test was} executed on a single core of \rev{a 2.8 GHz E5-2680v2 Intel Xeon Processor on the QB2 cluster of the Louisiana Optical Network Initiative (LONI)}. The code was compiled using the  \rev{GNU C++ compiler version 4.9.0}. The gravitational solution was generated, using opening angle $\theta = 0.2,0.3,...,1.0$, for
the original uncorrected D2000 method, the torque corrected method (using \eq{eq7}), and the torque corrected and optimized method (using \eq{eq8}). We refer to these three methods, respectively, as the 
``UC", ``TC", and ``TCO" variants.

\begin{figure}
\begin{subfigure}[b]{0.45\textwidth}
\includegraphics[scale=0.35]{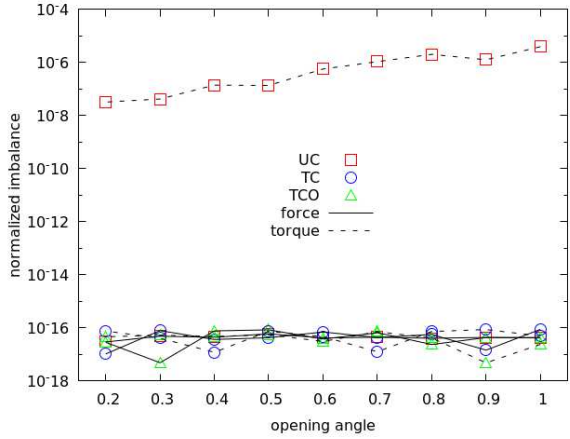} 
\caption{Net force imbalance}
\label{plot3}
\end{subfigure}
\quad
\begin{subfigure}[b]{0.45\textwidth}
\includegraphics[scale=0.35]{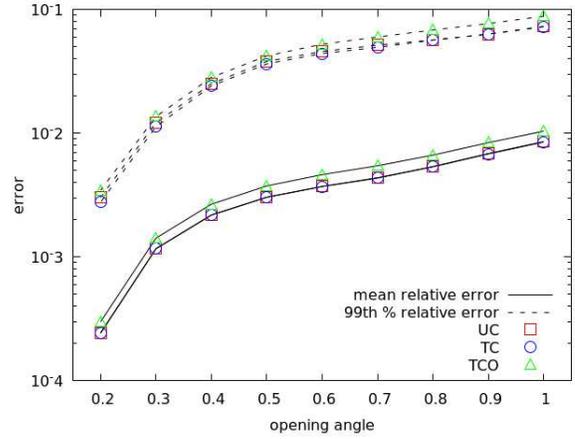} 
\caption{Mean relative force error }
\label{plot1}
\end{subfigure}
\quad
\begin{subfigure}[b]{0.45\textwidth}
\includegraphics[scale=0.35]{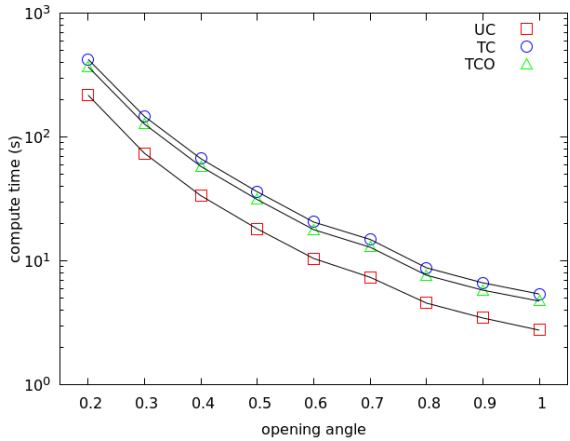}  
\caption{Compute time}
\label{plot2}
\end{subfigure}
\quad
\begin{subfigure}[b]{0.45\textwidth}
\includegraphics[scale=0.35]{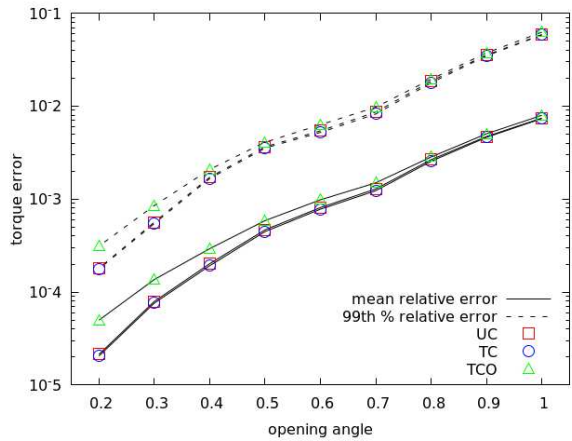}  
\caption{\rev{The mean relative torque error.} }
\label{plot4}
\end{subfigure}
\label{allplots}
\caption{\rev{Here we shows plots of (a) the force and torque imbalance , (b) the mean relative force error , (c) the compute time , and  (d) the mean relative torque error against opening angle, $\theta$, for each of the tested
FMM schemes. The original, uncorrected FMM of D2000 is refereed to as ``UC" (red squares), our torque corrected version is referred to as ``TC" (blue circles), and our optimized torque corrected version is referred to as ``TCO" (green triangles). }
}\end{figure}
The net force and torque balances were computed using the formula
\begin{equation}
\sqrt{\frac{\sum_d \left( \sum_n  \mu_n A_{n, d} \right)^2  }{\sum_d \left( \sum_n  \mu_n \left| A_{n, d} \right|  \right)^2 }},
\end{equation}
where  $A_{n, d}$ is the computed linear or angular acceleration, respectively, for the $n^\mathrm{th}$ particle and $d^\mathrm{th}$ dimension, and $\mu_n$ is the mass of the  $n^\mathrm{th}$ particle. As seen in \fig{plot3}, the TC and TCO variants preserve torque balance to many orders of magnitude greater precision than the UC variant, consistent with machine precision for double precision floating point arithmetic.
In all cases, net force balance is preserved to machine precision.
 \rev{In \fig{plot2} we plot the compute time for each variant. The TC variant takes approximately $2$ times the compute time of the UC variant, while the 
TCO variant takes $1.7$ times as long. The significant increase in computational time is due to the added expense of computing $D_{i j k l}$  (or ${D'}_{i j k l}$ )and $\vc{F_c}$. The computation of $\vc{F_c}$
adds $60$ terms ($20$ for each direction) to the $44$ terms present in \eq{eq2} when dipole moments are removed. The computation of $D_{i j k l}$ adds $83$ terms to the $27$ terms present in all
of the lower derivatives, while the optimized ${D'}_{i j k l}$ adds only $21$ more terms. } \rev{The} relative force error is defined as the average of 
\begin{equation}
 \epsilon_n = \frac{ \left | \vc{g_n} \ -  \vc{g_{PP,n}} \right | }{ \left | \vc{ g_{PP,n} } \right | } , 
\end{equation}
over all particles, where $\vc{g_{PP,n}}$ is the exact, directly computed force on the the $n^\mathrm{th}$ particle. \rev{The relative torque error is similarly defined using}
\begin{equation}
 \epsilon_n = \frac{ \left | \vc{X} \times \vc{g_n} \ -  \vc{X} \times \vc{g_{PP,n}} \right | }{ \left | \vc{X} \times \vc{ g_{PP,n} } \right | },
\end{equation}
\rev{where here $\vc{X}$ is the distance to the coordinate origin.We plot these errors in \fig{plot1} and \fig{plot4}. Both the force and torque errors are virtually identical between the
UC and TC variants. The torque correction in the TC variant does not result in a force or torque error higher than in the original scheme of the UC variant.  Both errors are higher for the TCO variant, therefore,}  for a given
error, the TCO variant requires a smaller $\theta$ than the TC variant, resulting in more interactions to compute.

\section{Discussion}
\label{discuss}
The method of \cite{D2000} conserves linear momentum to machine precision, but not angular momentum. We have presented two modifications to this method that each enable it to also conserve 
angular momentum to machine precision. This extra feature comes at computational expense,  \rev{requiring approximately
twice the compute time}.

Whether or not 
the extra
computational effort is worth the benefit of conserving angular momentum to machine precision will depend on the particular astrophysical system under investigation. One example of such a system would be 
a \rev{DWD} at the onset of stable mass transfer.  Past simulations of interacting DWDs have found angular momentum is artificially
 either added or removed from the system as the simulation progresses.
 \cite{MTF2002} found a normalized gain rate of $\sim 10^{-4} / \mathrm{orbit}$  for polytropic binaries of mass
 ratios $1.0$ and $0.8436$.
\cite{DRGR2011} found a $\sim  10^{-3} / \mathrm{orbit}$ normalized violation rate using SPH to simulate $84$ orbits of an interacting 0.8 $M_{\odot}$ accretor and 0.2 $M_{\odot}$ donor. 
These loss rates are sufficient that over many hundreds or more orbits, the violation of angular momentum conservation may cause systems that should be stable to become unstable  \rev{ (or vice versa) }.
One possible way to avoid this problem is to increase the resolution to the point the artificial angular momentum gain or loss rate is small compared 
to changes in the orbital and spin angular momenta of the system, however, it is difficult to determine what resolution is needed a priori. Increased resolution also comes at 
significant computational cost. The method described in this paper provides a 
remedy without increasing resolution.

 We also note that conservation of neither linear nor angular 
momentum holds when the time-stepping is not uniform throughout the entire domain. In practice many SPH and N-body codes use individual time-steps for particles or groups of particles, resulting in 
a faster computation speed \citep{AC1973}. In order to fully realize the benefits of the method presented here, one has to abandon individual time-stepping and the speed-up that comes with it. Another
benefit of individual time-stepping is that the
non-conservation of momentum is often used as a proxy for the measure of the force error. With exact conservation of momentum and angular momentum, this is no longer possible, necessitating the choice
of a different proxy. One possibility is to sum the magnitudes of the highest order expansion terms over the entire domain.


A higher order extension to this method is presented in the \S A. The method is also applicable for any Green's function that is solely a function of the scalar distance between points, 
such as softened gravitational potentials.

\section*{Acknowledgements}
We wish to acknowledge the support from the National Science Foundation through CREATIV grant AST- 1240655.

Portions of this research were conducted with high performance computing resources provided by Louisiana State University (http://www.hpc.lsu.edu).

Portions of this research were conducted with high performance computational resources provided by the Louisiana Optical Network Initiative (http://www.loni.org).

The author would like to thank Geoffrey C. Clayton, Patrick M. Motl, and Joel E. Tohline for their help with this article. The author also thanks the referee for help with the revision
process. The referee's insights and concerns enabled publication of an article of much greater quality than the original submission.

\appendix
\section{Appendix}

The $n^\mathrm{th}$ derivative of a Green's function dependent only on $R$, $D_{l_1 l_2 ... l_n}$, can be written 
\begin{equation}
D_{l_1 l_2 ... l_n}  := \frac{\partial}{\partial r_{l_1}} \frac{\partial}{\partial r_{l_2}} ... \frac{\partial}{\partial r_{l_n}} G\left(R\right).
\end{equation}
where the position vector, $\vc{r} := \vc{e_l} r_l$, is with respect to the coordinate center of the entire system.
The coordinate origin of cell $A$ is located at $\vc{X}$ and the coordinate origin of cell $B$ is located at $\vc{Y}$. The distance between the cells is $\vc{R} := \vc{Y} - \vc{X}$.
For the $i^\mathrm{th}$ particle in 
cell $A$, we define
$\vc{x_i}\left(\vc{r}\right) := \vc{r} - \vc{X}$,
and for the $j^\mathrm{th}$ particle in 
cell $B$, we define
$\vc{y_j}\left(\vc{r}\right) := \vc{r} - \vc{Y}$. The 
potential on the $i^\mathrm{th}$ particle of cell $A$ caused by all particles in cell $B$, expanded to order $P$, is
\begin{equation}
\label{d2002form}
\Phi_{i, B\rightarrow A} = \sum_{\vc{Y_j} \in \mathrm{cell B}}  \mu_j \sum_{m=0}^P \sum_{n=0}^{P-m}
\frac{\left(-1\right)^n}{n ! m !} x_{i,l_1} x_{i,l_2} ... x_{i,l_m} D_{l_1 l_2 ... l_m q_1 q_2 ... q_n}   y_{j,q_1} y_{j,q_2} ... y_{j,q_n}.
\end{equation}
 Equation (\ref{d2002form}) is Equation 3 from \cite{D2002} expressed in tensor notation. The potential on the $j^\mathrm{th}$ particle of cell $B$ caused by particles in cell $A$, expanded to order $P$, is
\begin{equation}
\Phi_{j, A\rightarrow B} = \sum_{\vc{X_i} \in \mathrm{cell A}}  \mu_i \sum_{m=0}^P \sum_{n=0}^{P-m}
\frac{\left(-1\right)^n}{n ! m !} x_{i,l_1} x_{i,l_2} ... x_{i,l_m} D_{l_1 l_2 ... l_m q_1 q_2 ... q_n}   y_{j,q_1} y_{j,q_2} ... y_{j,q_n}.
\end{equation}
Taking the negative of the derivative of $\Phi_{i, B\rightarrow A}$ with respect to $x_k$, we obtain the acceleration on the $i^\mathrm{th}$ particle of cell $A$, 
\begin{equation}
\vc{g_{i,B\rightarrow A}} = +\vc{e}_k  \sum_{\vc{Y_j} \in \mathrm{cell B}}  \mu_j \sum_{m=0}^{P-1} \sum_{n=0}^{P - 1 - m}
\frac{\left(-1\right)^{n}}{n ! m !} x_{i,l_1} x_{i,l_2} ... x_{i,l_m} D_{l_1 l_2 ... l_m k q_1 q_2 ... q_n}  y_{j,q_1} y_{j,q_2} ... y_{j,q_n}.
\end{equation}
Similarly, for the $j^\mathrm{th}$ particle of $B$, 
\begin{equation}
 \vc{g_{j,A\rightarrow B}} = -\vc{e}_k  \sum_{\vc{X_i} \in \mathrm{cell A}} \mu_i \sum_{m=0}^{P-1} \sum_{n=0}^{P-1-m}
\frac{\left(-1\right)^{n}}{n ! m !} x_{i,l_1} x_{i,l_2} ... x_{i,l_{m}} D_{l_1 l_2 ... l_m k q_1 q_2 ... q_n}  y_{j,q_1} y_{j,q_2} ... y_{j,q_n}.
\end{equation}
As expected, the sum of the forces over all particles is zero,
\begin{equation}
\sum_{\vc{X_i} \in \mathrm{cell A}}  \mu_i \vc{g_{i,B\rightarrow A}} + \sum_{\vc{Y_j} \in \mathrm{cell B}}  \mu_j\vc{g_{j,A\rightarrow B}} = 0.
\end{equation}
 The total torque, $\vc{\tau_{A B}}$, about the coordinate origin of cell $A$ is, 
\begin{equation}
\label{imbalance_deriv}
\begin{split}
\vc{\tau_{A B}} := \ & \sum_{\vc{Y_j} \in \mathrm{cell B}} \mu_j  \vc{R_j} \times \vc{g_{j, A\rightarrow B}} +
\sum_{\vc{X_i} \in \mathrm{cell A}} \mu_i \vc{x_i} \times \vc{g_{i, B\rightarrow A}} + \sum_{\vc{Y_j} \in \mathrm{cell B}} \mu_j \vc{y_j} \times \vc{g_{j, A\rightarrow B}}. 
\end{split}
\end{equation}
The total torque can be thought of as the sum of a bulk torque,
\begin{equation}
\label{bulkeq}
\begin{split}
\sum_{\vc{Y_j} \in \mathrm{cell B}} \mu_j \vc{R_j} \times \vc{g_{j, A\rightarrow B}} = & \\
                 \  \vc{e}_r \sum_{\vc{X_i} \in \mathrm{cell A}}  \sum_{\vc{Y_j} \in \mathrm{cell B}} &  \mu_i \mu_j \sum_{m=0}^{P-1} \sum_{n=0}^{P-1-m}    \frac{\left(-1\right)^n}    {n ! m !}  \epsilon_{ p k r } R_p x_{i,l_1} x_{i,l_2} ... x_{i,l_{m}} D_{l_1 l_2 ... l_m k q_1 q_2 ... q_n}  y_{j,q_1} y_{j,q_2} ... y_{j,q_n} \\
\end{split}
\end{equation}
and the spin torques of each cell,
\begin{equation}
\begin{split}
\sum_{\vc{X_i} \in \mathrm{cell A}} \mu_i \vc{x_i} \times \vc{g_{i, B\rightarrow A}} + \sum_{\vc{Y_j} \in \mathrm{cell B}} \mu_j \vc{y_j} \times \vc{g_{j, A\rightarrow B}} = & \\
                  \  \vc{e}_r  \sum_{\vc{X_i} \in \mathrm{cell A}}  \sum_{\vc{Y_j} \in \mathrm{cell B}} \mu_i \mu_j \sum_{m=0}^{P-1} \sum_{n=0}^{P - 1 - m} \frac{\left(-1\right)^{n+1}}{n ! m !} & \epsilon_{ p k r } x_{i,p} x_{i,l_1} x_{i,l_2} ... x_{i,l_m} D_{l_1 l_2 ... l_m k q_1 q_2 ... q_n}  y_{j,q_1} y_{j,q_2} ... y_{j,q_n} +\\
                   \ \vc{e}_r  \sum_{\vc{X_i} \in \mathrm{cell A}}  \sum_{\vc{Y_j} \in \mathrm{cell B}} \mu_i \mu_j  \sum_{m=0}^{P-1} \sum_{n=0}^{P-1-m}  \frac{\left(-1\right)^n}    {n ! m !}   & \epsilon_{ p k r } y_{j,p}  x_{i,l_1} x_{i,l_2} ... x_{i,l_{m}} D_{l_1 l_2 ... l_m k q_1 q_2 ... q_n}  y_{j,q_1} y_{j,q_2} ... y_{j,q_n}.
\end{split}
\end{equation}
Using fact that 
\begin{equation}
\label{spineq1}
\begin{split}
 R_p x_{i,l_1} x_{i,l_2} ... x_{i,l_{m}} & D_{l_1 l_2 ... l_m k q_1 q_2 ... q_n}  y_{j,q_1} y_{j,q_2} ... y_{j,q_n} =  \\ 
 & m x_{i,p} x_{i,l_1} x_{i,l_2} ... x_{i,l_{m-1}} D_{l_1 l_2 ... l_{m-1} k q_1 q_2 ... q_n}  y_{j,q_1} y_{j,q_2} ... y_{j,q_n} \ + \\
 & n y_{j,p} x_{i,l_1} x_{i,l_2} ... x_{i,l_{m}} D_{l_1 l_2 ... l_m k q_1 q_2 ... q_{n-1}}  y_{j,q_1} y_{j,q_2} ... y_{j,q_{n-1}},
\end{split}
\end{equation}
we can express the spin torques as
\begin{equation}
\label{spineq2}
\begin{split}
 \sum_{\vc{X_i} \in \mathrm{cell A}} \mu_i \vc{x_i} \times \vc{g_{i, B\rightarrow A}} + \sum_{\vc{Y_j} \in \mathrm{cell B}} \mu_j \vc{y_j} \times \vc{g_{j, A\rightarrow B}} = \\
                 \  -\vc{e}_r \sum_{\vc{X_i} \in \mathrm{cell A}}  \sum_{\vc{Y_j} \in \mathrm{cell B}} \mu_i \mu_j \sum_{m=0}^{P} \sum_{n=0}^{P-m}     \frac{\left(-1\right)^n}    {n ! m !} \epsilon_{ p k r } R_p x_{i,l_1} x_{i,l_2} ... x_{i,l_{m}} D_{l_1 l_2 ... l_m k q_1 q_2 ... q_n}  y_{j,q_1} y_{j,q_2} ... y_{j,q_n}.
\end{split}
\end{equation}
We see that the RHSs of  \eq{bulkeq} and \eq{spineq2} differ only in sign and the range of summation indices. The spin torques of expansion order $m$ are canceled by the bulk torque of expansion order $m+1$. The spin torques that result from the highest expansion order do not have 
a bulk torque to cancel them, resulting in a net torque. 
 Using \eq{imbalance_deriv},  \eq{bulkeq}, and \eq{spineq2}, we can express the net torque
\begin{equation}
\label{imbalance_eq}
\begin{split}
\vc{\tau_{A B}} = & \\
 \ \sum_{\vc{X_i} \in \mathrm{cell A}}  \sum_{\vc{Y_j} \in \mathrm{cell B}} & \mu_i \mu_j \vc{e}_r \sum_{n=0}^{P} \frac{\left(-1\right)^n}{n ! \left(P-n\right) !} 
\epsilon_{ p k r } R_p x_{i,l_1} x_{i,l_2} ... x_{i,l_{\left(P-n\right)}} D_{l_1 l_2 ... l_{\left(P-n\right)} k q_1 q_2 ... q_n}  y_{j,q_1} y_{j,q_2} ... y_{j,q_n}.
\end{split}
\end{equation}

 If we apply a constant corrective force, $\vc{F_c}$, to the particles in cell $A$, and  $-\vc{F_c}$ to the particles in cell $B$, the balance of force remains unaltered. The contribution
to the torque is
\begin{equation}
\label{tcorrect}
\vc{\tau_c} = \sum_{\vc{X_i} \in \mathrm{cell A}}  \sum_{\vc{Y_j} \in \mathrm{cell B}} \mu_i \mu_j \left\{-\vc{e}_r \epsilon_{p k q} R_p F_{c,k} +
 \vc{e}_r \epsilon_{p k q} x_{i,p} F_{c,k} - \vc{e}_r \epsilon_{p k q} y_{j,p} F_{c,k} \right\}.
\end{equation}
When the coordinate centers for cells $A$ and $B$ are coincident with the centers of mass for the respective cells, dipole moments vanish and the sum of corrective torques for
the last two terms on the RHS of \eq{tcorrect} vanish.
Comparing \eq{imbalance_eq} with the first term on the RHS of \eq{tcorrect}, we find that if we set
\begin{equation}
\vc{F_c} = \vc{e}_k \sum_{\vc{X_i} \in \mathrm{cell A}}  \sum_{\vc{Y_j} \in \mathrm{cell B}} \mu_i \mu_j \sum_{n=0}^{P} \frac{\left(-1\right)^n}{n ! \left(P-n\right) !} x_{i,l_1} x_{i,l_2} ... x_{i,l_{\left(P-n\right)}} D_{l_1 l_2 ... l_{\left(P-n\right)} k q_1 q_2 ... q_n}  y_{j,q_1} y_{j,q_2} ... y_{j,q_n},
\end{equation}
the sum of the original FMM torque and the corrective torque vanishes, 
\begin{equation}
\vc{\tau_{A B}} + \vc{\tau_c} = 0.
\end{equation}
Summing over all masses in each cell, the total corrective force, $\vc{F_C}$, is
\begin{equation}
\vc{F_c} = \vc{e}_k \sum_{n=0}^{P} \frac{\left(-1\right)^n}{n ! \left(P-n\right) !} M_{A, l_1 l_2 ... l_{\left(P-n\right)}} M_{B, q_1 q_2 ... q_n} D_{l_1 l_2 ... l_{\left(P-n\right)} q_1 q_2 ... q_n k}.
\end{equation}
Here we have defined the generalized moments for each cell,
\begin{equation}
M_{A, l_1 l_2 ... l_n} := \sum_{\vc{X_i} \in \mathrm{cell A}}  \mu_i x_{i,l_1} x_{i,l_2} ... x_{i,l_n}
\end{equation}
and
\begin{equation}
M_{B, q_1 q_2 ... q_m} := \sum_{\vc{Y_j} \in \mathrm{cell B}}  \mu_j y_{j,q_1} y_{i,q_2} ... y_{j,q_m}.
\end{equation}

Making the definition, 
\begin{equation}
{D'}_{k l_1 l_2 ... l_n} := {D}_{k l_1 l_2 ... l_n} -  \frac{R_k R_j}{R^2} {D}_{j l_1 l_2 ... l_n},
\end{equation}
we can define an alternative corrective force, 
\begin{equation}
\label{fceq}
\vc{{F'}_c} = \vc{e}_k \sum_{n=0}^{P} \frac{\left(-1\right)^n}{n ! \left(P-n\right) !} M_{A, l_1 l_2 ... l_{\left(P-n\right)}} M_{B, q_1 q_2 ... q_n} {D'}_{l_1 l_2 ... l_{\left(P-n\right)} q_1 q_2 ... q_n k}.
\end{equation}
This corrective force also results in a balanced torque. Depending on the choice of Green's function, \eq{fceq} may result in fewer terms to compute.

\end{document}